\documentclass[aps,prl,twocolumn,showpacs,groupedaddress]{revtex4}  

\usepackage{graphicx}  
\usepackage{dcolumn}   
\usepackage{bm}        
\usepackage{amssymb}   

\newcommand{\met}       {\mbox{$\not\!\!E_T$}}
\newcommand{\hatmet}    {{\mbox{$\not\!\!E'_T$}}}
\def\ppbar{p\overline{p}}            
\def\ttbar{t\overline{t}}            
\newcommand{\pythia}    {{\sc pythia}}
\newcommand{\alpgen}    {{\sc alpgen}}

\newcommand{\geant}    {{\sc geant}}

\newcommand{\cteqSixL}    {CTEQ6L1}

\begin{document}

\hspace{5.2in} \mbox{FERMILAB-PUB-08-267-E}

\title{$ZZ \rightarrow \ell^+\ell^-\nu\bar{\nu}$ production in $p\bar{p}$ collisions at $\sqrt{s}$= 1.96 TeV}
%
\author{V.M.~Abazov$^{36}$}
\author{B.~Abbott$^{75}$}
\author{M.~Abolins$^{65}$}
\author{B.S.~Acharya$^{29}$}
\author{M.~Adams$^{51}$}
\author{T.~Adams$^{49}$}
\author{E.~Aguilo$^{6}$}
\author{M.~Ahsan$^{59}$}
\author{G.D.~Alexeev$^{36}$}
\author{G.~Alkhazov$^{40}$}
% ...

\affiliation{\vspace{0.1 in}(The D\O\ Collaboration)\vspace{0.1 in}}
% ...
\affiliation{$^{6}$University of Alberta, Edmonton, Alberta, Canada,
                Simon Fraser University, Burnaby, British Columbia, Canada,
                York University, Toronto, Ontario, Canada, and
                McGill University, Montreal, Quebec, Canada}
% ...
\affiliation{$^{29}$Tata Institute of Fundamental Research, Mumbai, India}
% ...
\affiliation{$^{36}$Joint Institute for Nuclear Research, Dubna, Russia}
% ...
\affiliation{$^{40}$Petersburg Nuclear Physics Institute,
                St. Petersburg, Russia}
% ...
\affiliation{$^{49}$Florida State University, Tallahassee, Florida 32306, USA}
% ...
\affiliation{$^{51}$University of Illinois at Chicago,
                Chicago, Illinois 60607, USA}
% ...
\affiliation{$^{59}$Kansas State University, Manhattan, Kansas 66506, USA}
% ...
\affiliation{$^{65}$Michigan State University,
                East Lansing, Michigan 48824, USA}
% ...
\affiliation{$^{75}$University of Oklahoma, Norman, Oklahoma 73019, USA}
% ...
\date{August 2, 2008}

\begin{abstract}
We describe a search for $Z$ boson pair production in $p\bar{p}$ collisions at
$\sqrt{s}$= 1.96 TeV with the D0 detector at the Fermilab Tevatron Collider using a
data sample corresponding to an integrated luminosity of 2.7 fb$^{-1}$. 
Using the final state decay $ZZ\rightarrow \ell^+\ell^-\nu\bar{\nu}$
(where $\ell = e$ or $\mu$) we find a signal with
a 2.6 standard deviations significance (2.0 expected) corresponding to a
cross section of $\sigma(p\bar{p} 
\rightarrow ZZ + X) = 2.01 \pm 0.93 \mathrm{(stat.)}\pm 0.29 \mathrm{(sys.)}$~pb.
\end{abstract}

\maketitle 


\section{\label{sec:Intro}Introduction}
\begin{figure}[!Hhtb]
	\includegraphics[scale=0.6]{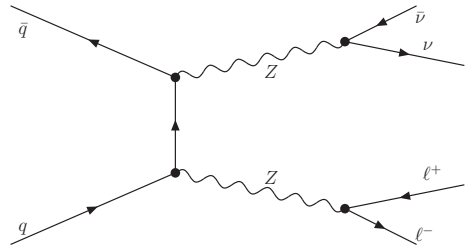}
	\caption{Leading order Feynman diagram for the process $ZZ
	\rightarrow \ell^+ \ell^- \nu \bar{\nu}$. \label{fig:feynman}}
\end{figure}

We report a search for $Z$ boson pair production in $p\bar{p}$
collisions in the mode where one $Z$ boson decays into two charged
leptons (either electrons or muons) and the other $Z$ boson decays into two
neutrinos (see Fig.~\ref{fig:feynman}). 
In the Standard Model (SM), $ZZ$ production is the double gauge boson process with the lowest cross section
and is the last remaining unobserved diboson process at the Tevatron,
aside from the expected associated production of the Higgs boson.
Observation of $ZZ$ production therefore represents an essential step
in Higgs boson searches in the $ZH$ and $WH$ channels with sensitivity at
the level of the expected SM cross sections.
Additionally, the $ZZ$ process forms a background to Higgs boson
searches, for example in the channels 
$ZH\rightarrow \ell^+\ell^- b\overline{b}$,  
$ZH\rightarrow \nu\overline{\nu} b\overline{b}$  
and $H \rightarrow W^+W^-\rightarrow\ell^+\nu\ell^-\overline{\nu}$.
Unlike the $WW$ and $WZ$ processes, there are no expected SM 
contributions from triple
gauge boson couplings involving two $Z$ bosons and a measurement of the $ZZ$
cross section represents a test for production of
this final state via anomalous couplings.

The process  $ZZ \rightarrow \ell^+\ell^-\nu\bar{\nu}$ has a branching ratio
six times larger than that for the other purely leptonic process
 $ZZ \rightarrow \ell^+\ell^-\ell'^+\ell'^-$.
After removing instrumental backgrounds, the dominant background in the $ZZ\rightarrow\ell^+\ell^-\nu\bar{\nu}$
search arises from the process
$WW\rightarrow\ell^+\nu\ell^-\bar{\nu}$, which produces the same final
state particles.
A kinematic discriminant against background from 
$WW\rightarrow\ell^+\nu\ell^-\bar{\nu}$ is employed.
In contrast, 
a search in the $ZZ\rightarrow\ell^+\ell^-\ell'^+\ell'^-$ channel benefits from
having no significant backgrounds from physics processes with the same
final state.

We select events containing an electron or muon pair with high invariant mass and significant
missing transverse momentum. After the initial event selection, the dominant
source of instrumental background to this signature arises from events
containing a leptonic $Z$ boson decay in which the apparent missing transverse momentum
arises from mismeasurement of the transverse momentum of either the charged
leptons or the hadronic recoil system.  We introduce a variable that is highly
discriminating against such instrumental background.

Although the $e^+e^- \rightarrow ZZ$ process has been observed at LEP~\cite{LEPZZ},
$ZZ$ production has not yet been observed at a hadron collider where
different physics processes are allowed and higher energies can be probed.
The D0 collaboration has previously performed a search for the process
$ZZ\rightarrow \ell^+\ell^-\ell'^+\ell'^-$ with
$\ell, \ell' = e\ \mathrm{or}\ \mu$~\cite{b-d0}, which set a limit on
the cross section of $\sigma(ZZ) < 4.4$ pb and also examined non SM
triple gauge boson couplings.
The CDF collaboration has recently produced a result using both the
$\ell^+\ell^-\ell'^+\ell'^-$ and the $\ell^+\ell^-\nu\bar{\nu}$
channels~\cite{b-cdf}, measuring the cross section to be $\sigma(ZZ) =
1.4^{+0.7}_{-0.6}$ pb.

The D0 detector \cite{nim1, nim2, nimmu} contains tracking, calorimeter and muon  
subdetector systems. 
Silicon microstrip tracking detectors (SMT) near the interaction point cover 
pseudorapidity $| \eta | < 3$ to provide tracking and vertexing information.
The SMT contains cylindrical barrel layers aligned with their axes parallel to the beams 
and disk segments.
The disks are perpendicular to the beam
axis, interleaved with, and extending beyond, the barrels.  
The central fiber tracker (CFT) surrounds the SMT, providing coverage to about
($| \eta | = 2$).
The CFT has eight concentric cylindrical layers of overlapped scintillating fibers
providing axial and stereo ($\pm 3^\circ$) measurements. 
A 2 T solenoid surrounds these tracking detectors.
Three uranium-liquid argon calorimeters measure particle energies. 
The central calorimeter (CC) covers $| \eta | < 1$, and two end calorimeters (EC) extend 
coverage to about $| \eta | = 4$. 
The calorimeter is highly segmented along the particle direction, 
with four electromagnetic (EM) and four to five hadronic sections, 
and transverse to the particle direction 
with typically $\Delta\eta = \Delta\phi = 0.1$, where $\phi $ is the azimuthal angle.
The calorimeters are supplemented with central and forward scintillating strip 
preshower detectors (CPS and FPS) located in front of the CC and EC.
Intercryostat detectors (ICD) provide added sampling in the region
$1.1 < |\eta | < 1.4$ where the CC and EC cryostat walls degrade the 
calorimeter energy resolution.
Muons are measured with stations which use scintillation counters and several layers of
tracking chambers over the range $| \eta | < 2$.
One such station is located just outside the calorimeters, with two more
 outside $1.8$~T iron toroidal magnets.
Scintillators surrounding the exiting beams allow determination of the luminosity.
A three level trigger system selects events for data logging at about 100 Hz.
The first level trigger (L1) is based on fast custom logic for several subdetectors and 
is capable of making decisions for each beam crossing.  
The second level trigger (L2) makes microprocessor 
based decisions using multi-detector information.  
The third level trigger (L3) uses fully digitized
outputs from all detectors to refine the decision and select events for offline processing.

\section{Data Set and Initial Event Selection}
The data for this analysis were collected with the D0 detector at the
Fermilab Tevatron $p\bar{p}$ Collider at a center-of-mass energy $\sqrt{s} =
1.96$ TeV.
An integrated luminosity of 2.7 fb$^{-1}$ is used after applying
data quality requirements.
The data are selected using a combination of single electron or single muon
triggers for the respective dilepton channels.

The data taking period prior to March 2006 is refered to as Run IIa, while IIb denotes the period after.
This division corresponds to the installation of an additional silicon vertex detector,
trigger upgrades, and a significant increase in the rate of delivered luminosity.

In each of the two channels we require that there be exactly two
oppositely charged leptons with transverse momentum $p_{T} > 15$~GeV and dilepton invariant
mass $70 < M_{\ell\ell} < 110$~GeV.
Electrons are required to be within the central ($|\eta| < 1.1$) or
forward ($1.5 < |\eta| < 2.5$) regions of the calorimeter. We do not
use electron candidates which point towards the transition region of the central
and forward cryostats. Electrons must pass
tight selection criteria on the energy close to them 
in $\Delta R$, where $\Delta R$ is the distance between two objects in
$(\eta,\phi)$ space, $\Delta R = \sqrt{(\delta\phi)^2 +
(\delta\eta)^2}$,
by requiring that
\begin{eqnarray*}
\frac{E_{\mathrm{tot}}(0.4) - E_{\mathrm{EM}}(0.2)}{E_{\mathrm{EM}}(0.2)} < 0.15,
\end{eqnarray*}
where $E_{\mathrm{tot}}(0.4)$ is the total energy within a cone of $\Delta R < 0.4$ and
$E_{EM}(0.2)$ is the EM energy within a cone of $\Delta R < 0.2$.
Additionally, a seven parameter multivariate discriminator compares the energy
deposited in each layer of the calorimeter and the total shower energy
to distributions determined from electron \geant{}
Monte Carlo (MC) simulations~\cite{b-geant}. This discriminator also uses the correlations between the various
energy distributions to ensure that the shower shape is consistent with
that produced by an electron.

Each muon is required to have an associated track in the central tracking system which has at least one hit
in the SMT and a distance of closest approach to the primary vertex in the
plane transverse to the beam of $|b|< 0.02$~cm. 
Furthermore, the muons must be isolated in both the calorimeter and
the tracker. For the former, a requirement is made that the sum of
calorimeter energies in cells within an annulus $ 0.1 <\Delta R <0.4$
around the muon track is smaller than 10\% of the muon $p_T$:
\begin{eqnarray*}
\Sigma^{cells}E_{T}(0.1<\Delta R<0.4)/p_{T}(\mu) < 0.1.
\end{eqnarray*}
For the latter, the sum of track $p_T$ within a cone $\Delta R < 0.5$
around the muon track (not included in the sum) must be smaller than 10\% of the muon $p_T$:
\begin{eqnarray*}
\Sigma^{track}p_{T}(\Delta R< 0.5)/ p_{T}(\mu) < 0.1.
\end{eqnarray*}

To suppress background from $WZ$ production, we veto events with one or
more leptons (e, $\mu$, or $\tau$) in addition to those forming the $Z$
candidate.  Additional lepton candidates must be separated by $\Delta R > 0.2$ from both
$Z$--candidate leptons. 
Electron candidates used in the veto must have $E_{T} > 5$~GeV and
either a central track match or satisfy shower shape requirements.  Muons
are rejected based on looser quality requirements than those from $Z$ decay.
Multi-prong hadronic taus are used to form the veto if they have been
identified using the standard D0 algorithms~\cite{b-d0tau}.  Finally, events
are vetoed if they have any isolated tracks with $p_{T} > 5$~GeV and a
separation distance between the track intercept with the beam line and the
primary vertex satisfying $|\Delta z| < 1$ cm.

Events with relatively large calorimeter activity are rejected by
vetoing on the presence of more than two jets in the detector. These
jets are reconstructed using the Runn IIa cone algorithm~\cite{jetcone}
with a radius of 0.5 and
must satisfy $\Delta R$(jet, lepton)$ > 0.3$ and jet $E_{T} > 15$~GeV.  This
requirement significantly reduces background from $t\overline{t}$ production.

Missing transverse energy ($\met$) is the magnitude of the vector sum of transverse
energy above a
set threshold in the calorimeter cells, corrected for the jets in the
event. At the dilepton selection stage, we do not make a requirement on $\met$.

\section{\label{sec:sim}Background and Signal Prediction}

Background yields were estimated using a combination of control data
samples and MC simulation.  The primary background after the initial
selection is inclusive $Z/\gamma^*\rightarrow \ell^+\ell^-$ production.  After making the
final selection described later, the dominant background is 
$W^+W^-\rightarrow\ell^+\nu\ell^-\overline{\nu}$ events.  Additional
backgrounds include $\ttbar$ production, $WZ$ production and $W\gamma$ or
$W$+jets events in which the $\gamma$ or jet is misidentified as an
electron.

The $WW$, $WZ$, $Z/\gamma^*$, and $t\bar{t}$ backgrounds are estimated based on simulations using the 
\pythia~\cite{b-pythia} event generator, with the 
leading order \cteqSixL~\cite{b-cteq} parameterization used for the parton distribution functions
(PDFs).  We pass the simulated events through a detailed D0 detector simulation
based on \geant{} and reconstruct them using the same
software program used to reconstruct the collider data.  The
$Z/\gamma^*$ MC events are assigned a weight as a function of generator
level $p_T$, to match the $p_T$ spectrum observed in unfolded data~\cite{zpt}.
Randomly triggered collider data events are added to the simulated \pythia\
events.
These data events are taken at various instantaneous luminosities to
provide a more accurate modeling of effects related to the presence of
additional $p\bar{p}$ interactions and
detector noise.  We also apply corrections for trigger efficiency,
reconstruction efficiency and identification efficiency.  The
corrections are derived from comparisons of control data samples with
simulation.

The $W\gamma$ background is estimated from a calculation of the
Next to Leading Order (NLO)
 production cross section~\cite{b-baur}, which we use to normalize
events generated by \pythia.  The probability for a photon to be
misidentified as an electron is measured in a
$Z\rightarrow\ell^+\ell^-\gamma$ control data sample.  This probability is
then applied to the $W\gamma$ yield predicted from simulation to determine the
contribution to our selected sample in which the photon is mistakenly 
reconstructed as the second electron.

The kinematic distributions of the $W$+jet events are
determined from simulation based on \alpgen~\cite{b-alpgen}.
The overall yield from $W$+jet events is determined from
data. 
The probabilities that an electron or a jet which satisfy looser
selection requirements will also pass our candidate selection in data
are measured by solving  a set of linear equations
involving these
probabilities, the number of candidate events, and the number of
events in which one of the requirements on one of the
candidates has been loosened.
The solution to this set of linear equations is used to determine the number of
$W$+jet events in the final sample.
The fraction of jets misidentified as electrons is ~$2 \times 10^{-3}$, and the probability that a jet fakes a muon is ~$4 \times 10^{-4}$
(averaging over Run IIa and Run IIb).

The relative normalization of the background sources determined from simulation
is taken from ratios of NLO cross sections, and the absolute normalization of
the total background is then determined by matching the observed yield under the $Z$ dilepton mass peak to the
predicted background sum after applying the basic $Z$ selection and
extra--activity event veto.  
We choose this normalization method because
inclusive $Z$ production dominates our signal by four orders of magnitude, and this approach allows
cancellation of multiplicative scale uncertainties and systematics
related to the modeling of the selection efficiencies.
The normalization factor agrees with that obtained using the
integrated luminosity to within the associated 6\% uncertainty.

Signal events were generated using \pythia\ with the \cteqSixL\
PDFs, and the signal event samples were corrected for the same detector effects as
the background samples.

\def\ual{\hat{a_l}}
\def\uat{\hat{a_t}}

\section{\label{sec:Recoil}Verification of Missing Transverse Momentum}
The basic event signature for this analysis is a high mass pair of
charged leptons from the decay of a $Z$ boson, produced in association with
significant missing transverse momentum $\met$ arising from the neutrinos produced
in the decay of a second $Z$ boson.  Substantial
background comes from single inclusive $Z$ production in which
mismeasurement results in a mistakenly large $\met$ value.  Because the cross
section times branching ratios for inclusive $Z$ production and $ZZ$ signal
differ by a factor of more than $10^4$, stringent selection criteria against
inclusive $Z$ production are required. We present a novel approach to this
challenge.  In particular, we do not attempt to make an unbiased or accurate
estimate of the missing transverse momentum in the candidate events.  Rather,
our approach is to construct a variable $\hatmet$ which is a representation of
the minimum $\met$ feasible given the measurement uncertainties of the leptons
and the hadronic recoil.  Thus, this is not intended to be the best estimator
of true $\met$, but rather to be robust against reconstruction mistakes.  This
approach is inspired by the OPAL collaboration which used a similar variable to search
in final states similar to that of this analysis~\cite{OPAL}.

The $\hatmet$ variable is constructed in five steps.  
\renewcommand{\labelenumi}{\roman{enumi}.}
\begin{enumerate}
\item The first step is the computation of a reference axis chosen
such that effects from leptonic resolution occur dominantly along this
axis, and the decomposition of the dilepton system transverse momentum
into components parallel and perpendicular to this reference axis.
\item The second step is determination of a recoil variable based on
measured calorimeter jet or total calorimeter activity.  In events
with no significant neutrino energy or mismeasurement, the quantities
calculated in the first two steps should be approximately balanced.
\item The third step is the calculation of a correction based on
recoil track $p_T$ for tracks which are well separated from the
candidate leptons and calorimeter jets. 
\item The fourth step is the computation of a correction term
accounting for lepton transverse momentum measurement uncertainties.  
\item The final step is a combination of the quantities
computed in the first four steps into the $\hatmet$ variable. 
\end{enumerate}

\subsection{Decomposition}
\begin{figure}[!Hhtb]
  \centering
  \includegraphics[angle=0,scale=0.4]{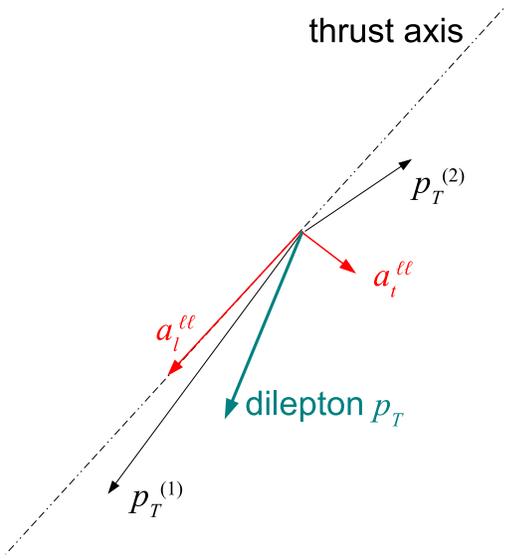}
  \caption{Representation of the transverse plane of the event
    and of the decomposition of the dilepton transverse momentum
    along the thrust axis.\label{fig:thrust_alat}}
\end{figure}

In the first step, to minimize the sensitivity to mismeasurement of
the $p_T$ of the individual leptons, the $\vec{p}_T$ of the lepton pair is
decomposed into two components, one of which is almost insensitive to lepton
$p_T$ resolution for $Z$ candidates with moderate values of transverse
momentum.  This decomposition is achieved as
follows.  In the transverse plane a dilepton thrust axis is defined (see
Fig.~\ref{fig:thrust_alat}).
This axis maximizes the scalar sum of the projections of the $p_{T}$ of the two
leptons onto the axis.  It is defined as
$$
\vec{t}  =  \vec{p}_{T}^{\ (1)} - \vec{p}_{T}^{\ (2)},
$$ 
where $\vec{p}_{T}^{\ (1)}$ and $\vec{p}_{T}^{\ (2)}$ are the transverse momenta of
the higher and lower $p_{T}$ leptons respectively.  

We then define two unit vectors $\ual$ and $\uat$ which are parallel and
perpendicular to the thrust axis.
For the rest of this paper, a subscript $l$ denotes the component in the
$\ual$ directio and a subscript $t$ denotes the component of a vector
in the $\uat$ direction.

\begin{figure}[!Hhtb]
  \centering
  \includegraphics[scale=0.21]{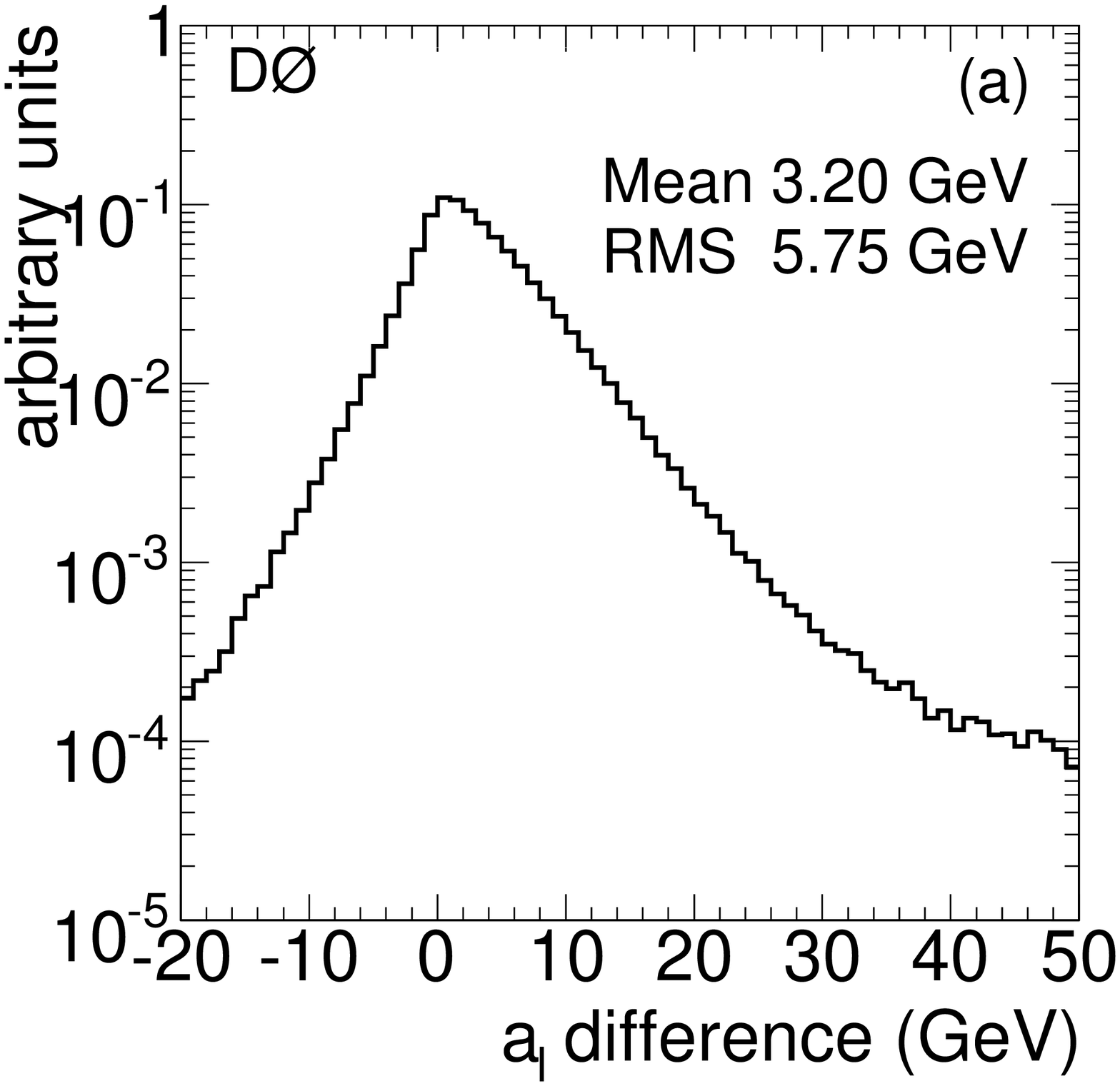}
  \includegraphics[scale=0.21]{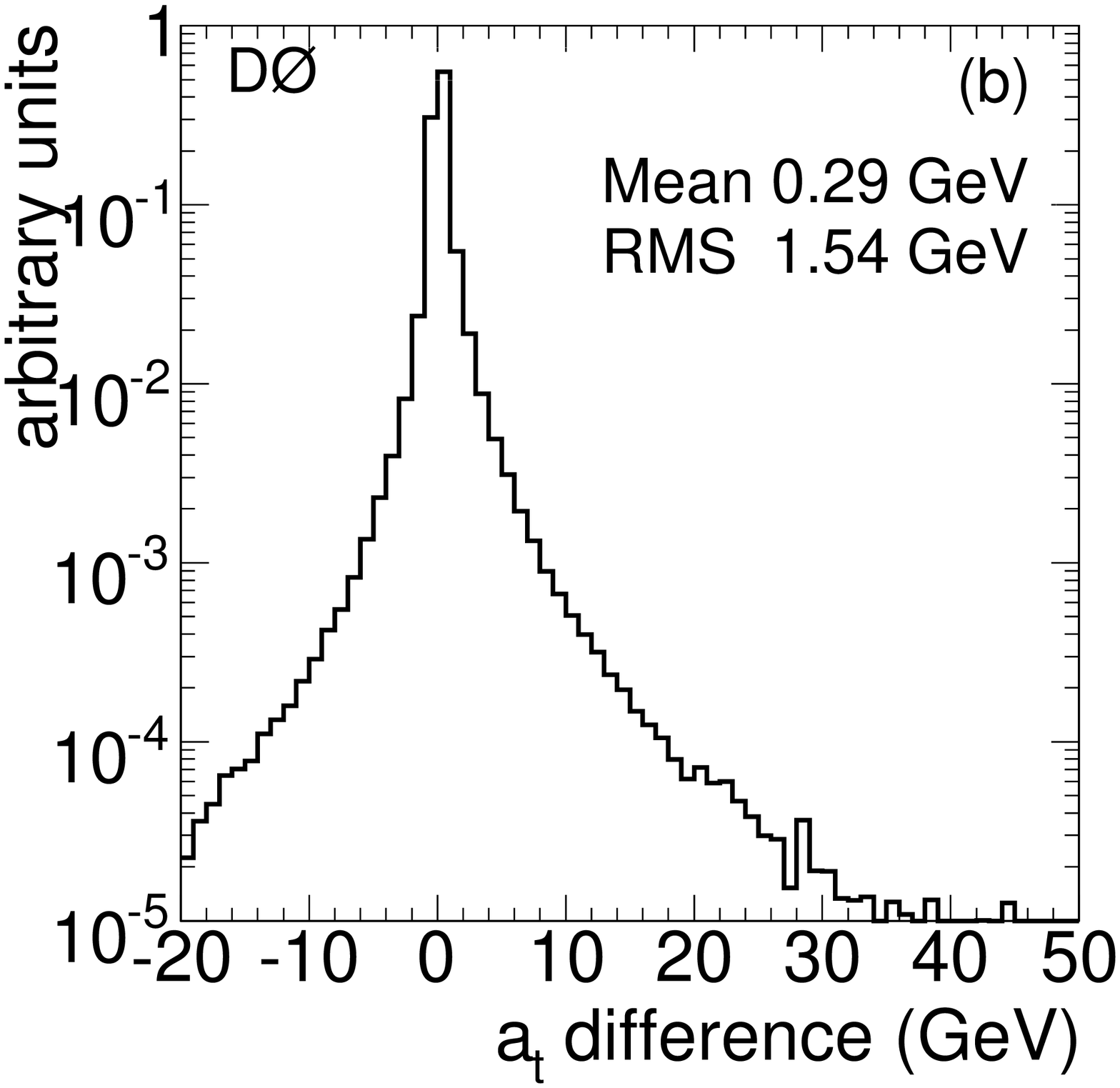}
  \caption{The difference between measured and true dilepton $p_T$ in 
    simulated $Z\rightarrow \mu\mu$ events projected along the (a) $\ual$ and 
    (b) $\uat$ axes.\label{fig:decompdemo}}
\end{figure}

The dilepton system transverse momentum is decomposed into
components parallel to $\ual$ ($a_{l}^{\ell\ell}$) and perpendicular to $\ual$
($a_{t}^{\ell\ell}$).  These are given by
\begin{eqnarray*}
  a_{t}^{\ell\ell} & = & \vec{p}_T^{\ \ell\ell} \cdot \uat \\
  a_{l}^{\ell\ell} & = & \vec{p}_T^{\ \ell\ell} \cdot \ual,
\end{eqnarray*}
in which $\vec{p}_T^{\ \ell\ell} \equiv \vec{p}_T^{\ (1)} + \vec{p}_T^{\ (2)}$
is the dilepton system transverse momentum.  The resolutions of the two
components are shown in Fig.~\ref{fig:decompdemo} from $Z\rightarrow
\mu\mu$ (MC) generated events. Resolution effects are more pronounced in
the $Z\rightarrow\mu\mu$ channel than in the $Z\rightarrow ee$.  As
seen, the lepton momentum resolution
effects are significantly more pronounced in the $\ual$ direction than
in the $\uat$ direction.

The decomposition is performed only for events in which $\Delta
\phi^{\ell\ell} > \pi/2$, where $\Delta \phi^{\ell\ell}$ is the angle
between the two charged leptons in the transverse plane.  For the case
$\Delta\phi^{\ell\ell} \leq \pi/2$ the direction of the dilepton transverse
momentum $\vec{p}_T^{\ \ell\ell}$ is used to define $\uat$, and all components in
the $\ual$ direction are set to zero.

\subsection{Calorimeter Recoil Activity}
The second step in the process uses calorimeter energy to assess whether or not
it might generate apparent, though false, $\met$.  Two measures of net
calorimeter transverse energy are considered: (a) a vector sum of the $E_T$
of selected reconstructed jets and (b) the uncorrected missing transverse
energy $\met$ for the event from the calorimeter. When
computing the jet $\vec{p}_T$ sum, we consider only those jets whose $p_T$ is
in the direction opposite to the dilepton system for each of the $\ual$ and
$\uat$ directions.  That is, if $\vec{p}_T^{\ jet}\cdot\ual<0$, then this amount is
added to the $\ual$ correction, and if $\vec{p}_T^{\ jet}\cdot\uat<0$, it is
added to the $\uat$ correction. 

The calorimeter activity correction is then defined using either the jets or
the uncorrected $\met$.  We chose the one with the largest projected magnitude
along each of the two axes in the hemisphere opposite to the dilepton
pair.  Additionally, we allow for the
possibility that only some of the true recoiling energy is underestimated by
multiplying the observed energy by two. The calorimeter correction is thus defined as
\begin{eqnarray*}
  \delta a_t^{cal} & = & 2\times\mathrm{min}(\Sigma \vec{p_T}^{jets}\cdot\uat,-\vec{\met}\cdot\uat,0) \\
  \delta a_l^{cal} & = & 2\times\mathrm{min}(\Sigma \vec{p_T}^{jets}\cdot\ual,-\vec{\met}\cdot\ual,0)
\end{eqnarray*}
in which a jet is used in the $\ual(\uat)$ component sum only if it
satisfies $\vec{p}_T\cdot\ual<0\ (\vec{p}_T\cdot\uat<0)$.  The presence of the
zero term in the $\mathrm{min}$--function ensures that the calorimeter activity
correction is used only if it {\it decreases} the apparent value of
$a_{t}^{\ell\ell}$ and/or $a_{l}^{\ell\ell}$.  In this way we try to minimize
the possibility that a well-balanced event acquires an apparently significant
net missing momentum due to calorimeter noise or a jet with a grossly
overestimated energy.

\subsection{Recoiling Tracks}
The third step identifies events in which the recoil activity is not observed
in the calorimeter as jets.  We consider tracks that are $\Delta R > 0.5$ away from all
calorimeter jets, $\Delta R > 0.5$ from the candidate leptons, have a fit satisfying ${\chi^{2}}/{\mathrm{NDF}} < 4.0$, and $p_T
> 0.5$~GeV and use these to build track jets using a cone algorithm.  A track
jet is seeded by the highest $p_T$ track not yet associated with any track jet.
All tracks which are within $\Delta R<0.5$ of the seed track and are not yet
associated with a track jet are added to the current track jet.  The track jet
transverse momentum $p_T^{tjet}$ is the vector sum of the $p_T$ values of all
tracks forming the track jet.  This is repeated until no unused tracks
outside the calorimeter jet cones remain.
A track--based correction is then defined as
\begin{eqnarray*}
  \delta a_t^{trk} & = & (\Sigma\vec{p}_T^{\ tjet})\cdot\uat\\
  \delta a_l^{trk} & = & (\Sigma\vec{p}_T^{\ tjet})\cdot \ual.
\end{eqnarray*}
As with the calorimeter jets, a track jet is included in the correction for
the $\ual(\uat)$ direction only if it satisfies $\vec{p}_T^{\ tjet}\cdot\ual<0\ 
(\vec{p}_T^{\ tjet}\cdot\uat<0)$.

\subsection{Lepton $p_T$ Uncertainty}
In the fourth step of the algorithm, corrections $\delta a_{t}^{\ell\ell}$ and
$\delta a_{l}^{\ell\ell}$ arising from the uncertainties in the lepton transverse
momenta are derived. 
The basic approach taken is to fluctuate the lepton transverse momenta
by one standard deviation of their uncertainty so as to minimize, separately, the
 $\ual$ and $\uat$ components of the dilepton $p_T^{\ell\ell}$.
The transverse component, $a_t^{\ell\ell}$, is minimized by decreasing the transverse momenta of
both leptons to give the modified quantity:
$$
  a_{t}^{\ell\ell'}  =  \vec{p}_T^{\ \ell\ell'} \cdot \uat'. 
$$
  Here $\vec{p}_T^{\ \ell\ell'}$ and $\uat'$ correspond, respectively,
  to $\vec{p}_T^{\ \ell\ell}$ and $\uat$,
redefined using $\vec{p}_T^{\ (1)}\times(1-\sigma_1)$ and
$\vec{p}_T^{\ (2)}\times (1-\sigma_2)$ in place of the unscaled
  quantities.
The uncertainty is then simply given by:
$$
  \delta a_{t}^{\ell\ell} = a_{t}^{\ell\ell'} - a_{t}^{\ell\ell}.
$$
The longitudinal component, $a_l^{\ell\ell}$, is minimized by decreasing $\vec{p}_T^{\ (1)}$ and
increasing $\vec{p}_T^{\ (2)}$ using their fractional uncertainties $\sigma_1$ and $\sigma_2$: 
$$
  \delta a_{l}^{\ell\ell} = (-\sigma_1\vec{p}_T^{\ (1)} +  \sigma_2\vec{p}_T^{\ (2)})\cdot\ual 
$$
If the fractional uncertainty on either of the
lepton transverse momenta is larger than unity, then the fractional
uncertainties on both $a_t^{\ell\ell}$ and $a_l^{\ell\ell}$ are set to unity.

Electrons falling at calorimeter module boundaries of the
central calorimeter require special treatment as their calculated uncertainties do not
reflect the probability for such electrons to have very significantly
underestimated energies.  To account for this, if the lower $p_{T}$ electron is
within a central calorimeter module boundary, then the fractional
uncertainty on $a_l^{\ell\ell}$ is set to unity.

\subsection{Combination}
In the fifth and final step, the variable $\hatmet$ is computed from the quantities
calculated in the previous steps.  We compute components:
\begin{eqnarray*}
 a_t & = & a_t^{\ell\ell} + \delta a_t^{cal} + k'\times \delta a_t^{trk} + k\times \delta a_{t}^{\ell\ell}\\
 a_l & = & a_l^{\ell\ell} + \delta a_l^{cal} + k'\times \delta a_l^{trk}  + k\times \delta a_{l}^{\ell\ell},
\end{eqnarray*}
where $k$ and $k'$ are constants defined below.
Recall that by construction the $\delta a_i$ (where $i = t$ or $l$) terms are always zero or negative while
$a_i^{\ell\ell}$ is positive.

For events with significant transverse energy
from neutrinos and no mismeasurements, the $a_i$ variables are large and
positive. 
If  $a_i \leq 0$, then there is no significant missing transverse
momentum along direction $i$ and that component is ignored in the
subsequent analysis by setting:  
\begin{eqnarray*}
a_t' & = & \mathrm{max}(a_t,0),\\
a_l' & = & \mathrm{max}(a_l,0).
\end{eqnarray*}

The final discriminating variable is then calculated as a weighted
quadrature sum of the two components:
\begin{eqnarray*}
 \hatmet = \sqrt{a_l'^2 + (1.5a_t')^2}.
\end{eqnarray*}
By construction $\hatmet$ is less that $\met$.  The factor of $1.5$ is used with the $\uat$ component
 to give extra weight to the better--measured direction.
As mentioned earlier, if $\Delta\phi^{\ell\ell}<\pi/2$, then instead of using
the thrust axis, the reference axis direction $\uat$ is simply the
$\vec{p}_T^{\ \ell\ell}$ direction, and the $\ual$ components are ignored. 

The values $k$ and $k'$ were optimized by applying a loose cut in
$\hatmet$ (such that the background in the sample is
dominated by $Z\rightarrow\ell^+\ell^-$ events) and maximizing $S/\sqrt{B}$,
where $S$ is the number of signal events and $B$ is the number of
background events. 
The chosen values for
dielectron events are $k=2.2$ and $k'=2.5$. For dimuon events, $k=k'=1.5$.

\begin{figure}[!Hhtb]
  \centering
  \includegraphics[scale=0.35]{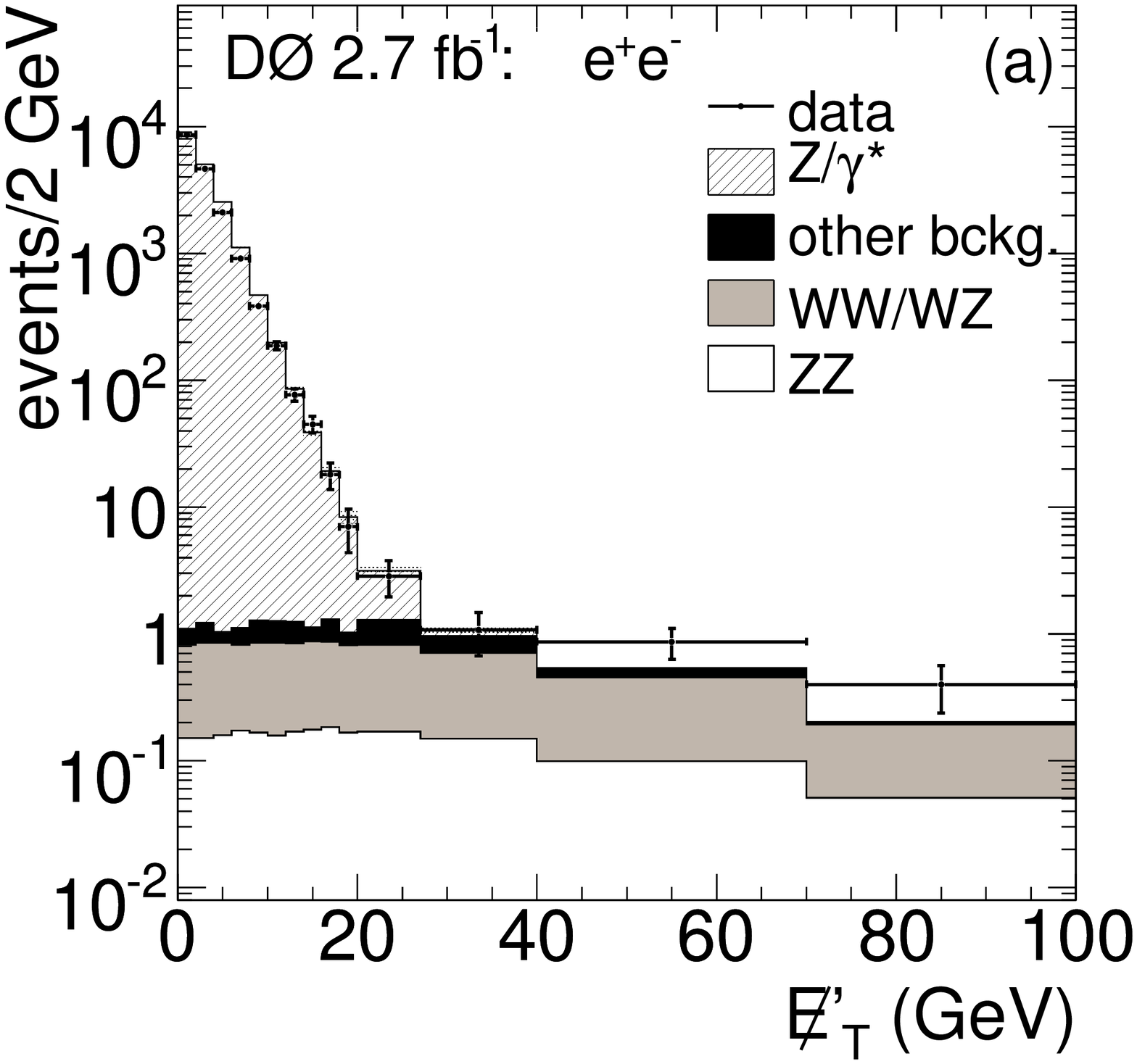}
  \includegraphics[scale=0.35]{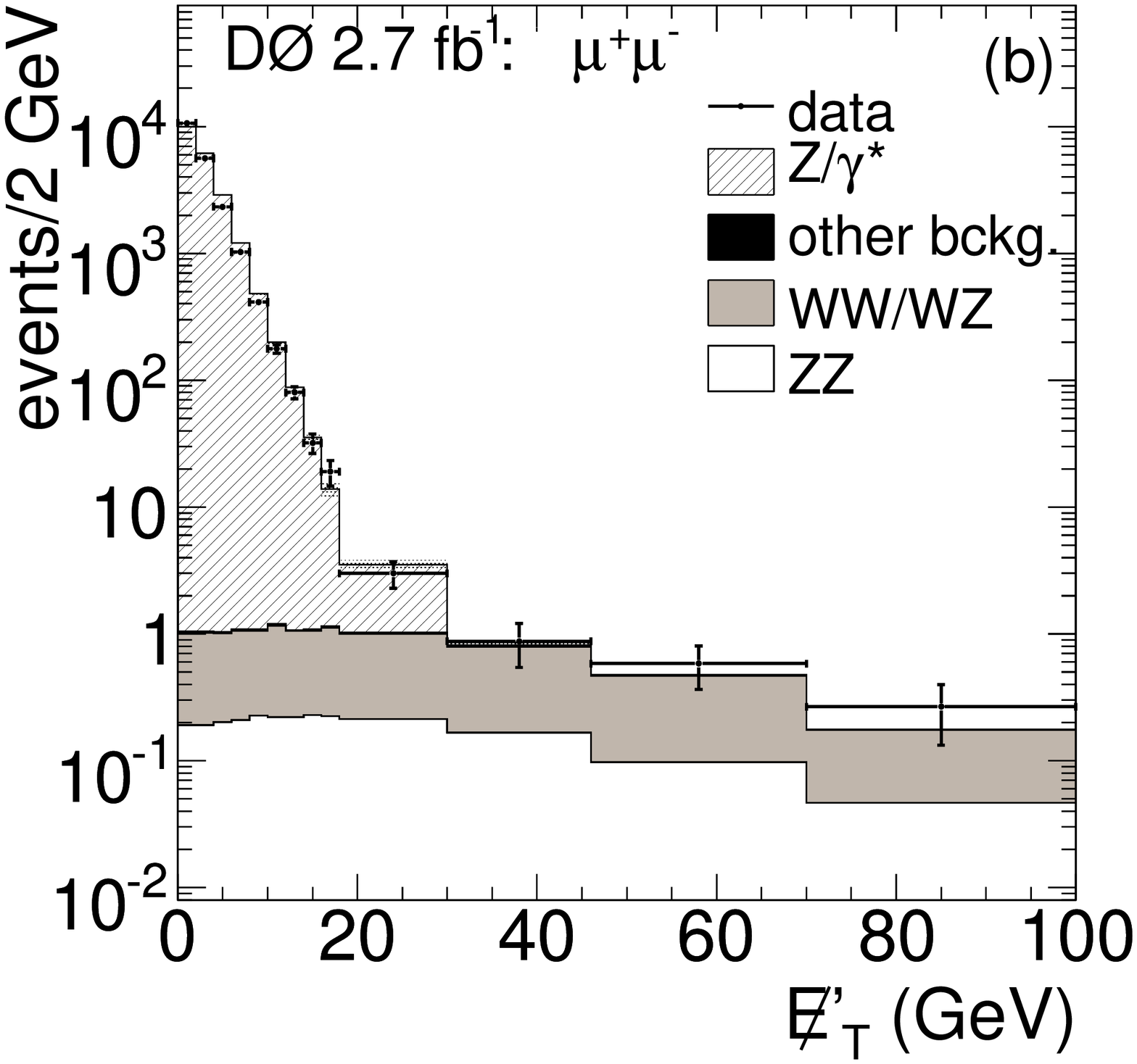}
  \caption{The $\hatmet$ distributions for dielectron (a) and dimuon
  (b) events. \label{fig:metReduced}}
\end{figure}

The power of the variable $\hatmet$ is displayed in Figure~\ref{fig:metReduced}, which shows the distribution of  $\hatmet$,
 for the dielectron and dimuon channels separately.

\begin{figure}[!Hhtb]
  \centering
  \includegraphics[scale=0.2]{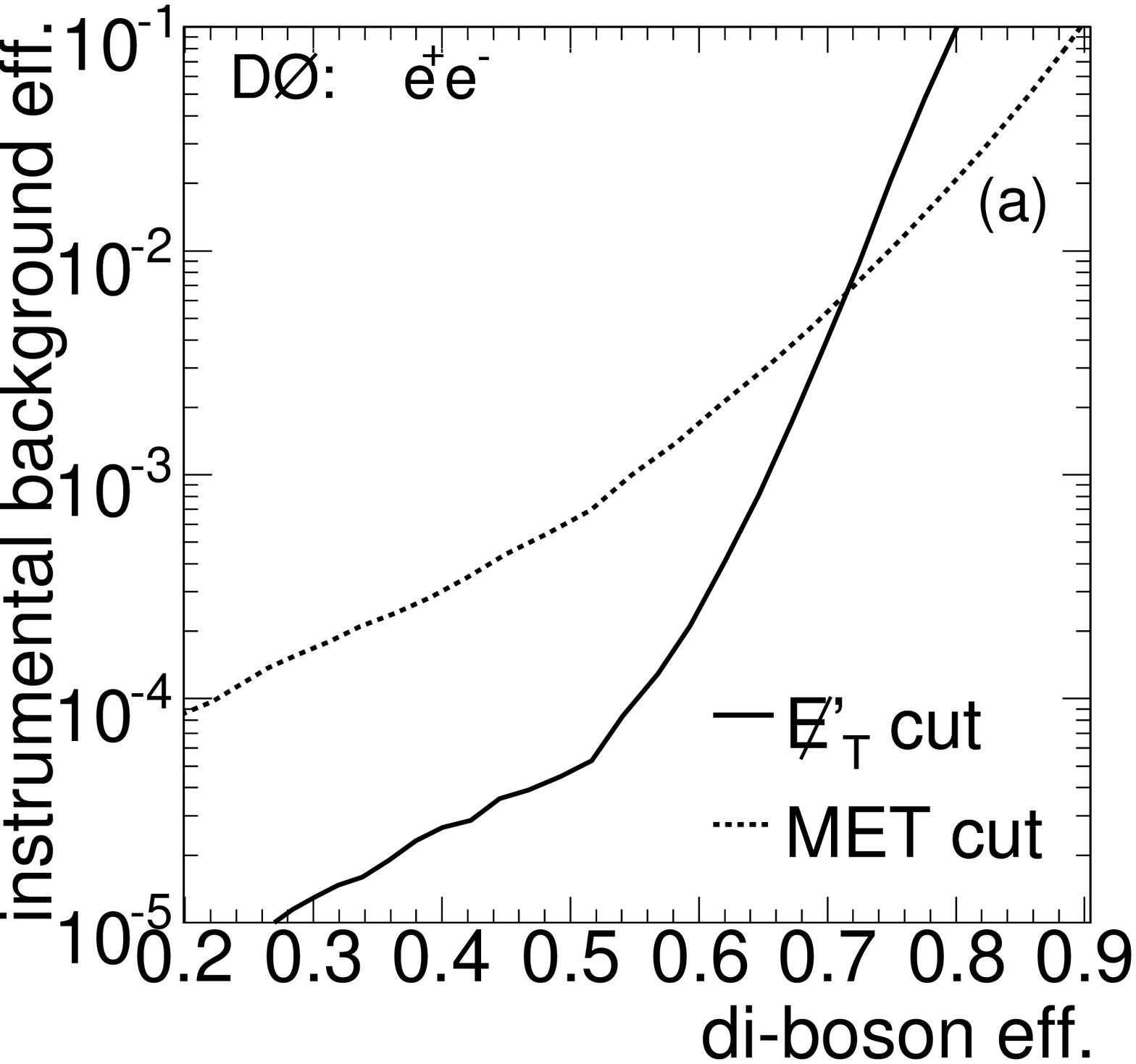}
  \includegraphics[scale=0.2]{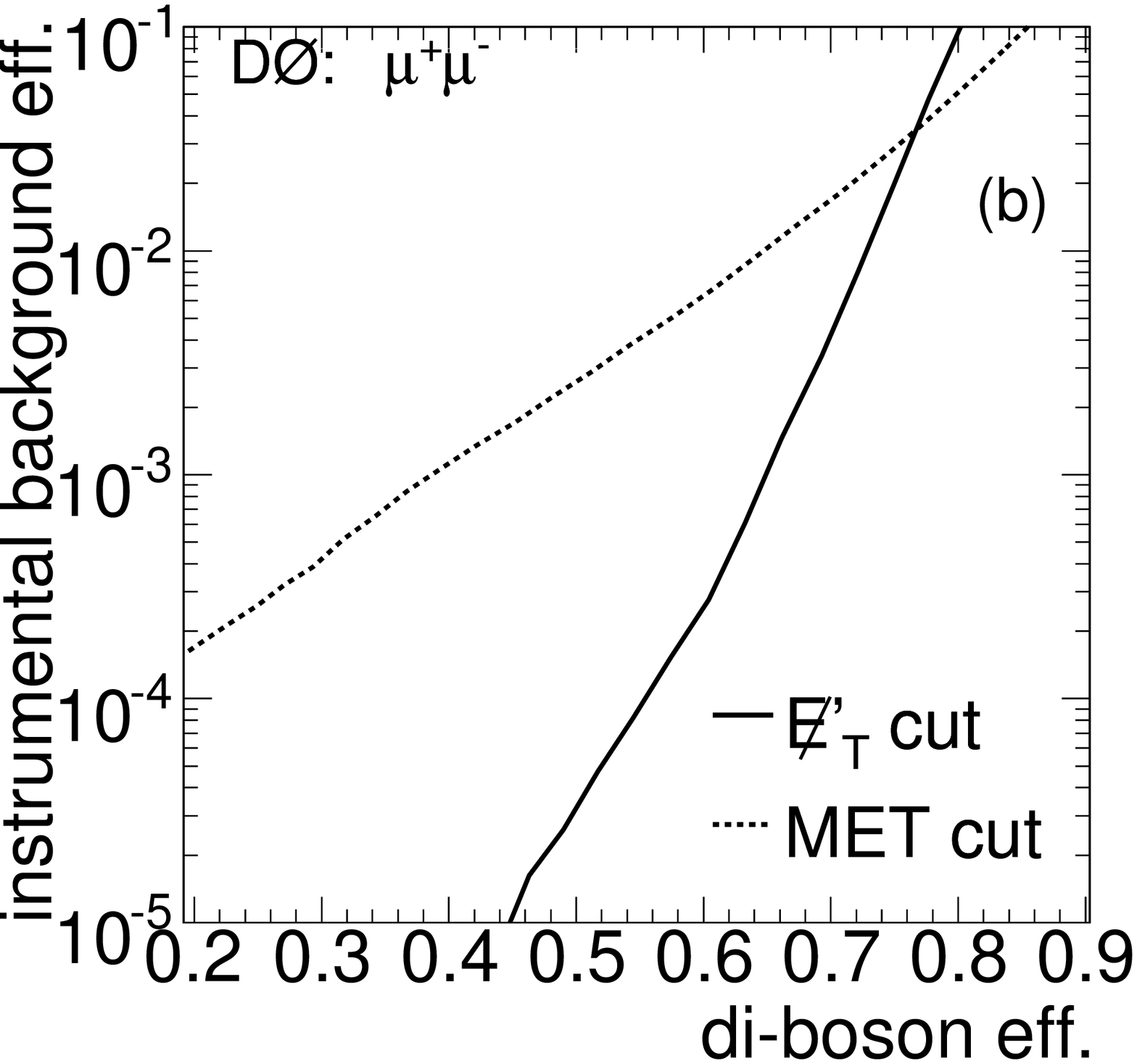}
  \caption{The efficiency of $WW$, $WZ$, and $ZZ$ events vs single $Z$
	   events using the $\met$ (labeled as MET in the figure) and $\hatmet$ variables in the
	   dielectron (a) and dimuon (b) channels. \label{fig:rejection}}
\end{figure}

The  separation of sources with true $\met$ (e.g. $WW$ and $ZZ$) from
those without, especially inclusive $Z$ production, is clearly visible.
The rejection of events from single $Z$ boson production using this method and a simple $\met$ cut is shown in Figure~\ref{fig:rejection}.

\section{\label{sec:Yields}Final Selection, Likelihood and Yields}
In addition to the initial selection requirements, events selected for
further analysis must satisfy,
\begin{eqnarray*}
  \hatmet & > & 27\ \mathrm{GeV,\ dielectron,\ IIa}, \\
  \hatmet & > & 30\ \mathrm{GeV,\ dimuon,\ IIa}, \\
  \hatmet & > & 27\ \mathrm{GeV,\ dielectron,\ IIb}, \\
  \hatmet & > & 35\ \mathrm{GeV,\ dimuon,\ IIb}.
\end{eqnarray*}
The values of these requirements were chosen so as to optimize  the expected significance of a $ZZ$
observation in the four individual analysis channels,
assuming the SM cross sections for signal and backgrounds.  
The effect of systematic uncertainties,
as described below, were used in the significance calculation for the optimization.
Tables~\ref{tab:cutFlowDiem} and \ref{tab:cutFlowDimu} show the predicted and
observed yields after the initial selection and after the $\hatmet$ selection
for the dielectron and dimuon channels respectively.  The requirement on
$\hatmet$ reduces the predicted instrumental background yields well below those
for our signal and remaining physics backgrounds. 
In the dielectron channel, we observe 8 events (8.9 expected) in the
IIa data, with another 20 events (10.7 expected) in the IIb data. Of these,
we expect 1.8 and 2.3 to be signal events respectively. In the
dimuon channel, we observe 10 events (7.0 expected) in the IIa data and 5 events (7.3
expected) in the IIb data. Here, we expect 1.7 signal events in each
data set.

\def\hmyOO{\hphantom{$\pm$ 0.00}}
\begin{table}[hbt]
\caption{Number of predicted dielectron events and yield observed in data after the dilepton
        selection and after the requirement on $\hatmet$. The uncertainties in the final column are
        statistical only.  If not present, the statistical uncertainty
        is negligible.}
\centering 
\label{tab:cutFlowDiem}
\begin{ruledtabular}
\begin{tabular}{l c c c c}
\hline
Sample  & dilepton selection & $\hatmet$ requirement  \\
\hline
$Z\rightarrow \ell^+ \ell^-$            & $1.18\times10^5$    & ~0.5 $\pm$ 0.2 \\
$Z\rightarrow\tau^+\tau^-$              & 48.3      & 0.35 \hmyOO \\
$W$+Jets                              & 18.2      & ~2.7 $\pm$ 0.4 \\
$t\bar{t}$                          & 16.4      & 0.34 \hmyOO \\
$WW$                                  & 28.0      & 10.6 $\pm$ 0.1 \\
$WZ$                                  & 19.2      & 1.08 \hmyOO \\
$W\gamma$                          &  2.0      & 0.03 \hmyOO \\
Predicted Background                & $1.19\times10^5$    & 15.6 $\pm$ 0.4 \\
\hline
$ZZ\rightarrow \ell^+\ell^-\ell'^+\ell'^-$  & 2.9       & 0.02 \hmyOO \\
$ZZ\rightarrow \ell^+\ell^-\nu\bar{\nu}$     & 8.9       & 4.03 \hmyOO \\
Predicted Total                     & $1.19\times10^5$    & 19.6 $\pm$ 0.4 \\
\hline
Data                                & 118,850    & 28 \\
\hline
\end{tabular}
\end{ruledtabular}
\end{table}

\begin{figure}[!Hhtb]
  \centering
  \includegraphics[scale=0.2]{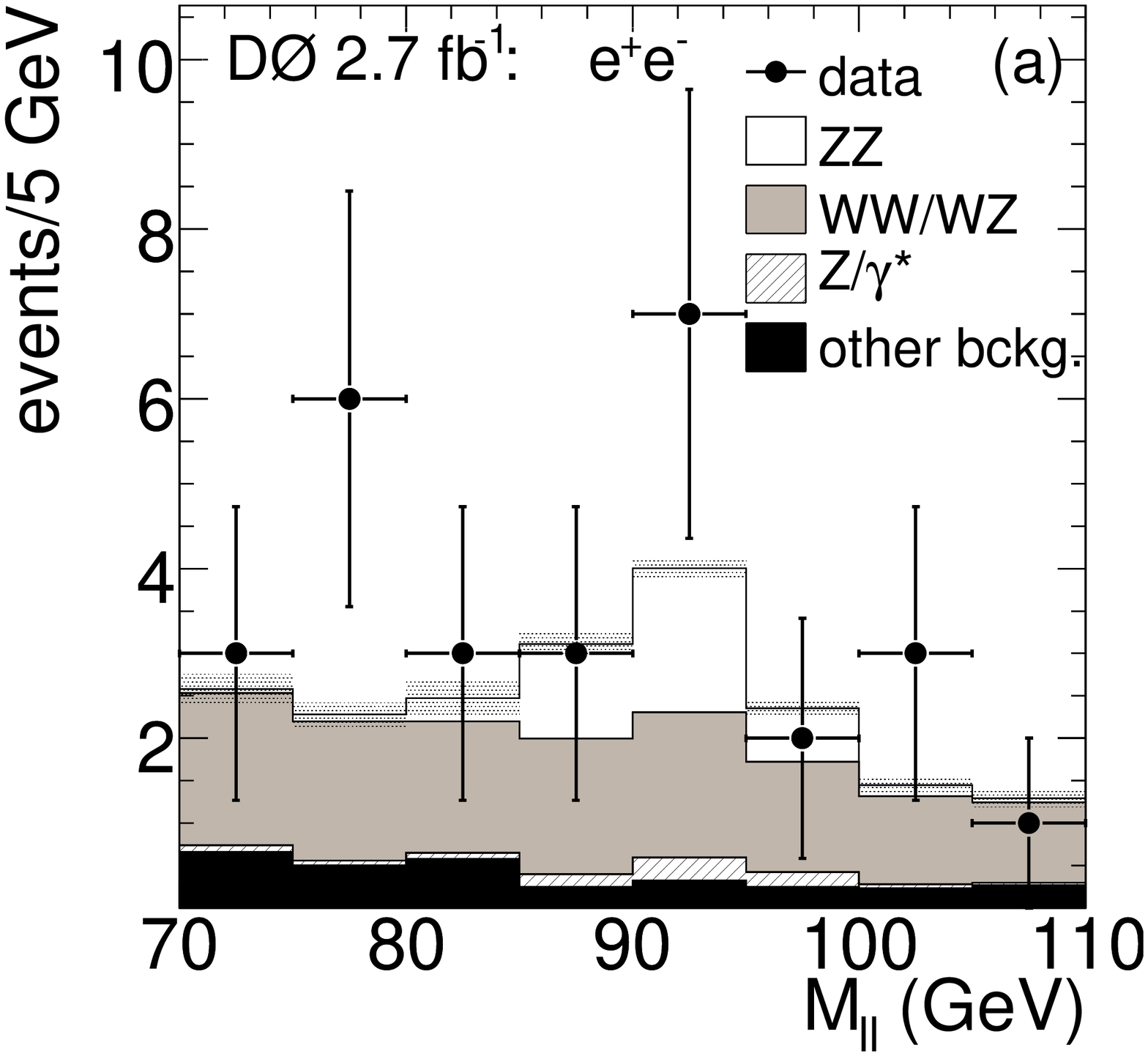}
  \includegraphics[scale=0.2]{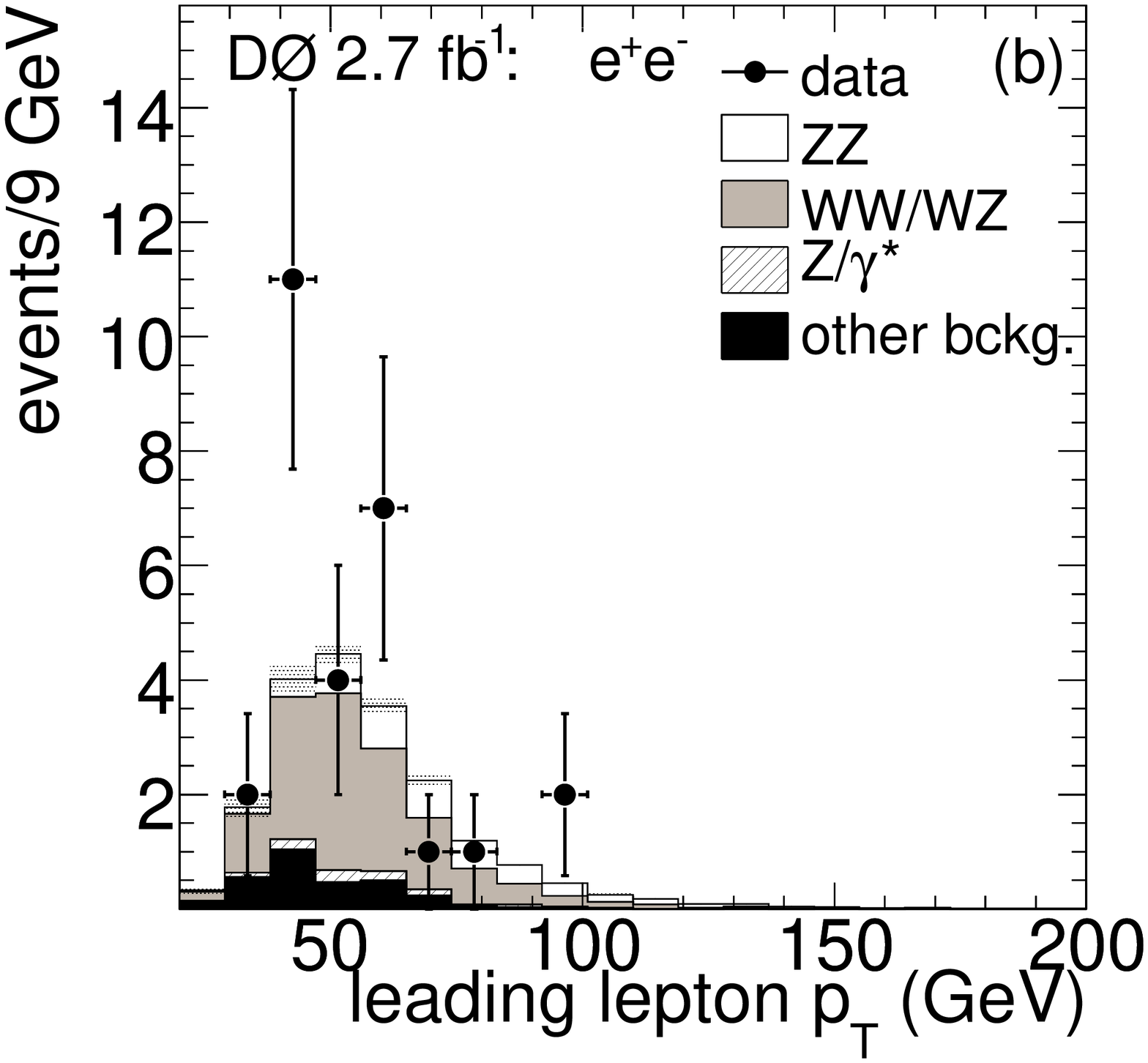}
  \includegraphics[scale=0.2]{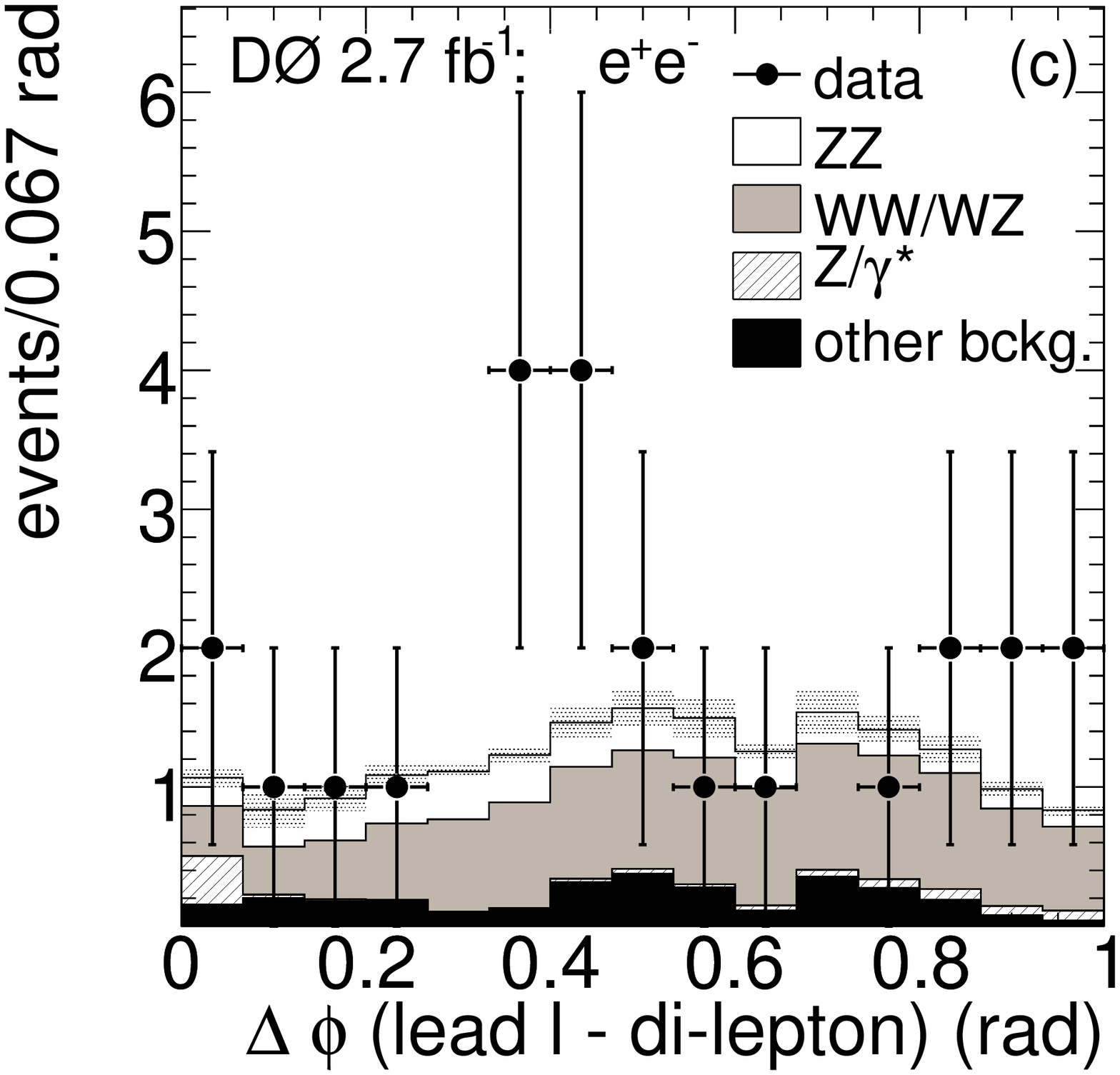}
  \includegraphics[scale=0.2]{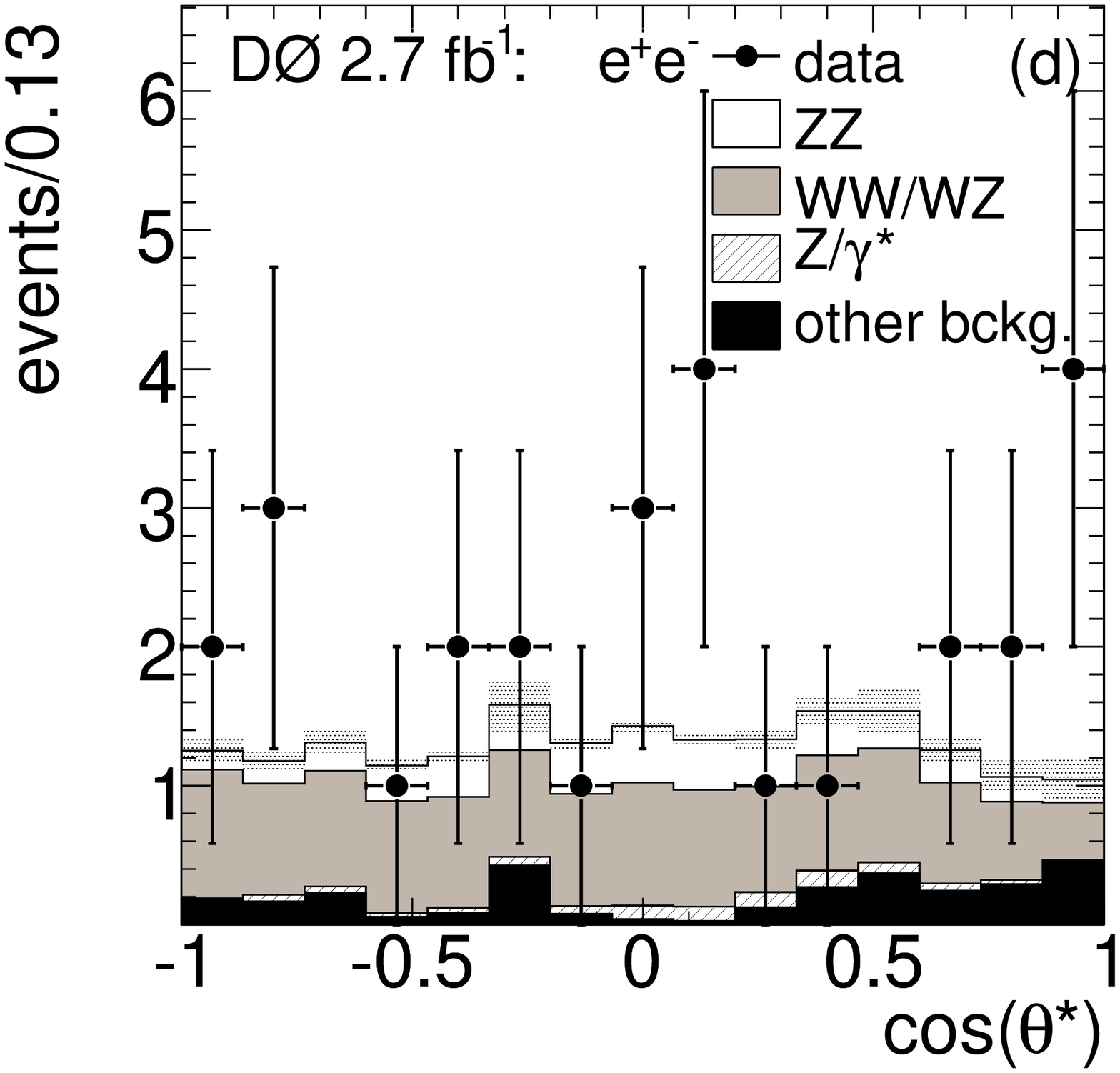}
  \caption{Distribution in the dielectron channel of the input variables of the likelihood
    discriminant for data and MC.
    Invariant mass of the dilepton system (a), $p_{T}$ of the 
    leading lepton (b), the opening angle between the
    lead lepton and the dilepton system (c), and the cosine of 
    the scattering angle of the negative
    lepton in the dilepton rest frame (d).
    \label{fig:LLInput_diem}}
\end{figure}

\begin{figure}[!Hhtb]
  \centering
  \includegraphics[scale=0.2]{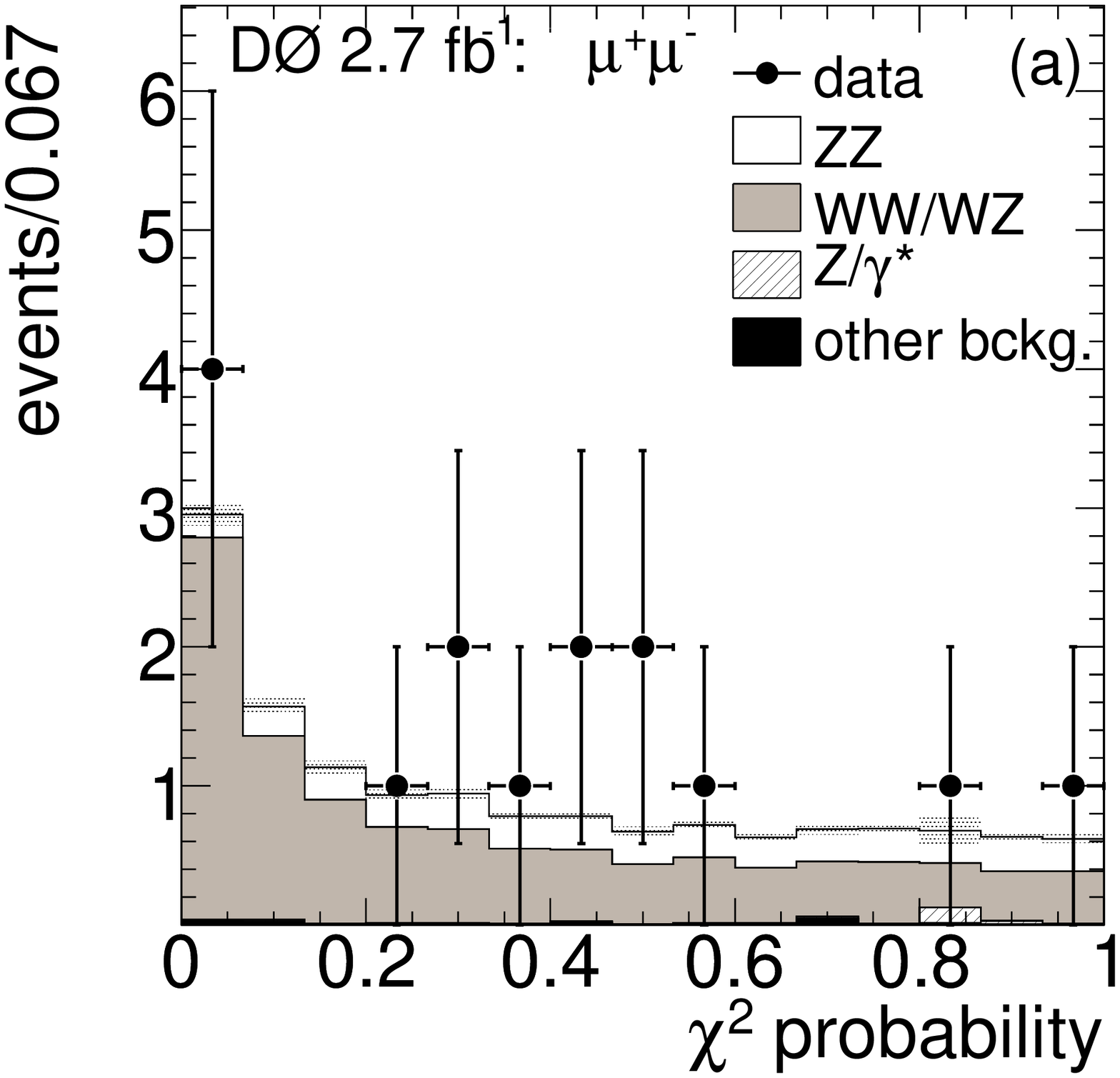}
  \includegraphics[scale=0.2]{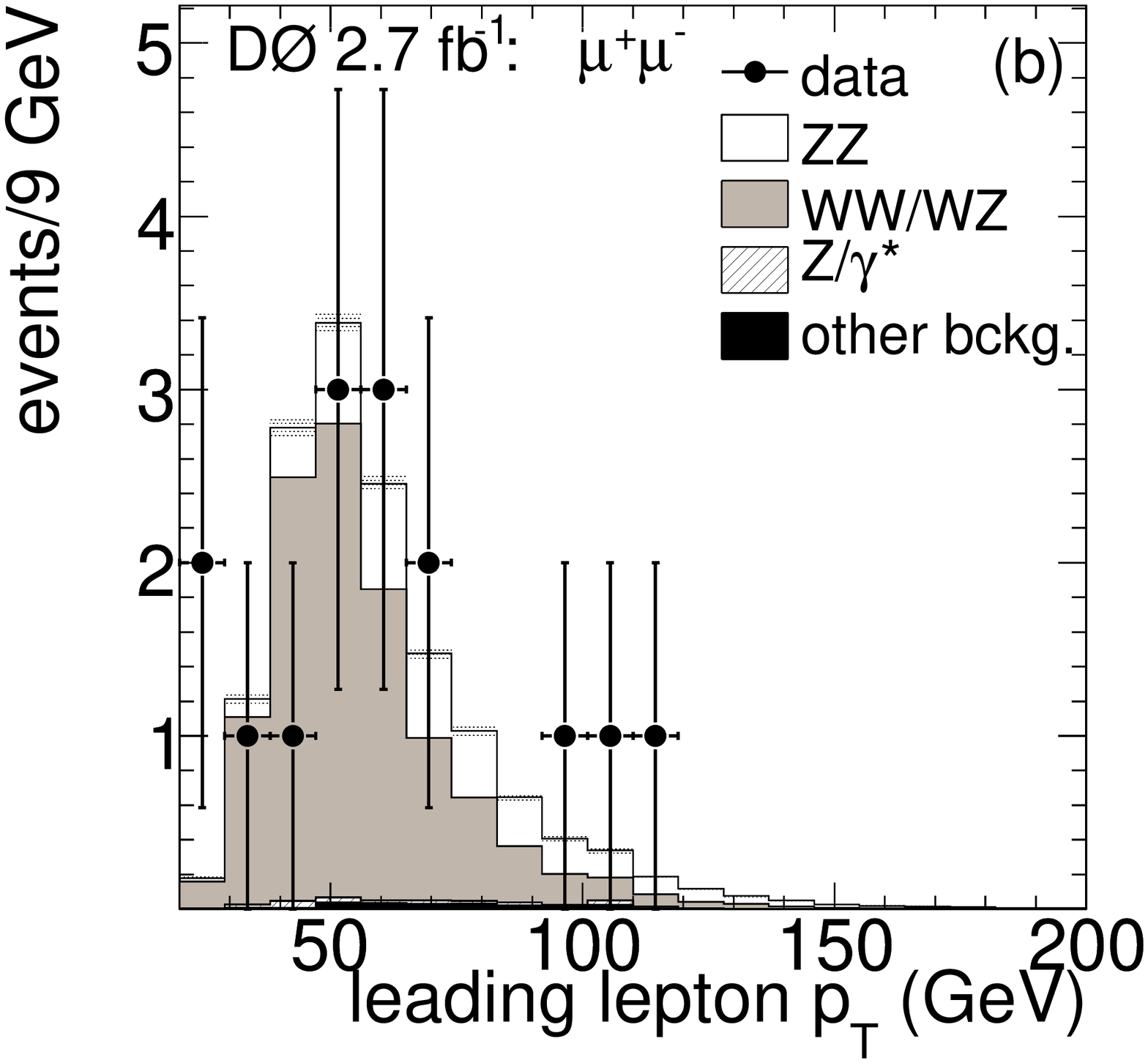}
  \includegraphics[scale=0.2]{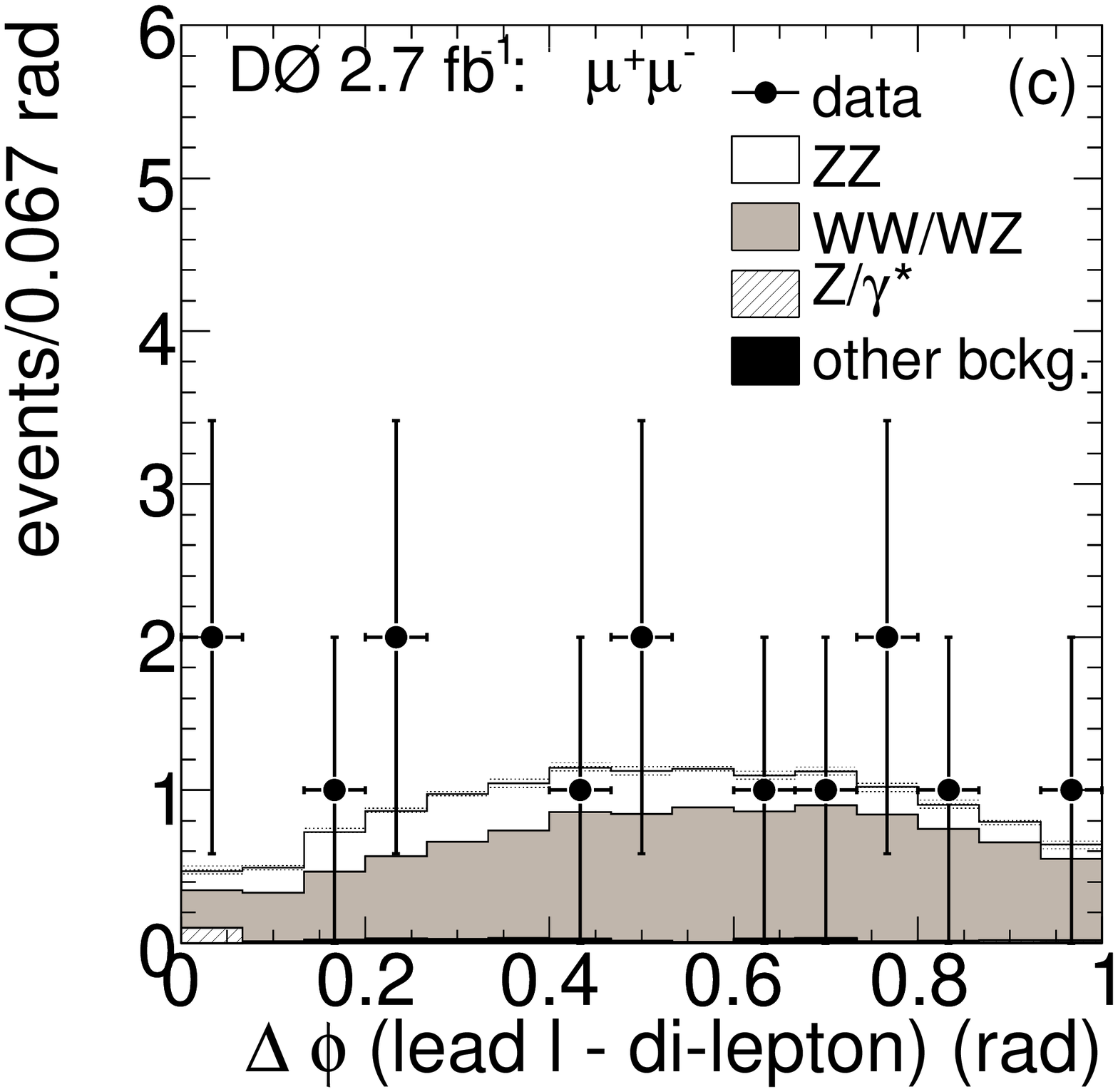}
  \includegraphics[scale=0.2]{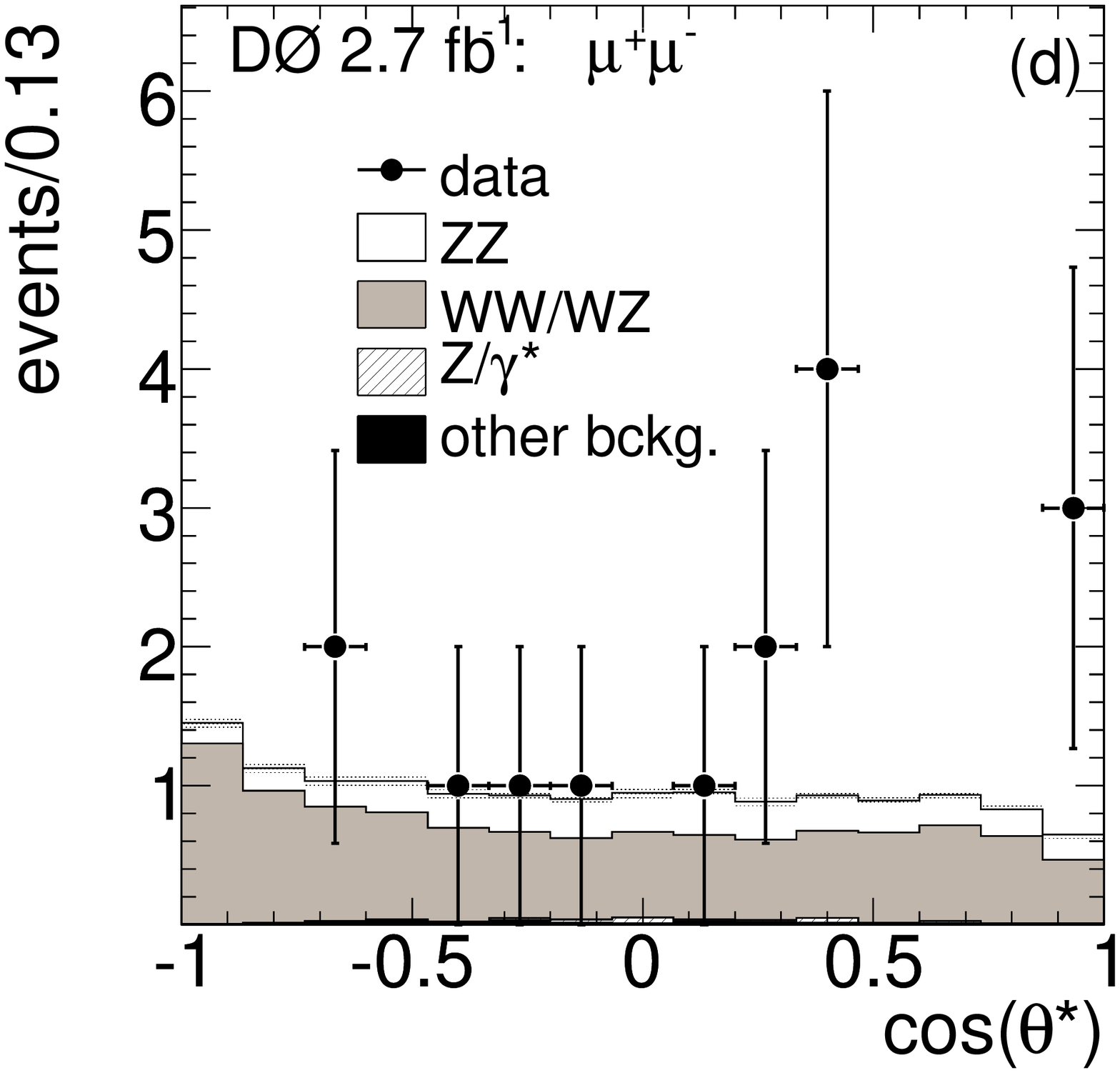}
 \caption{Distribution in the dimuon channel of the input variables of the likelihood
    discriminant for data and MC.
    $\chi^{2}$ probability for the
    kinematic fit to the dilepton mass (a), $p_{T}$ of the 
    leading lepton (b), the opening angle between the
    lead lepton and the dilepton system(c), and the cosine of
    the scattering angle of the negative
    lepton in the dilepton rest frame (d).
    \label{fig:LLInput_dimu}}
\end{figure}

\begin{figure}[!Hhtb]
  \centering
  \includegraphics[scale=0.2]{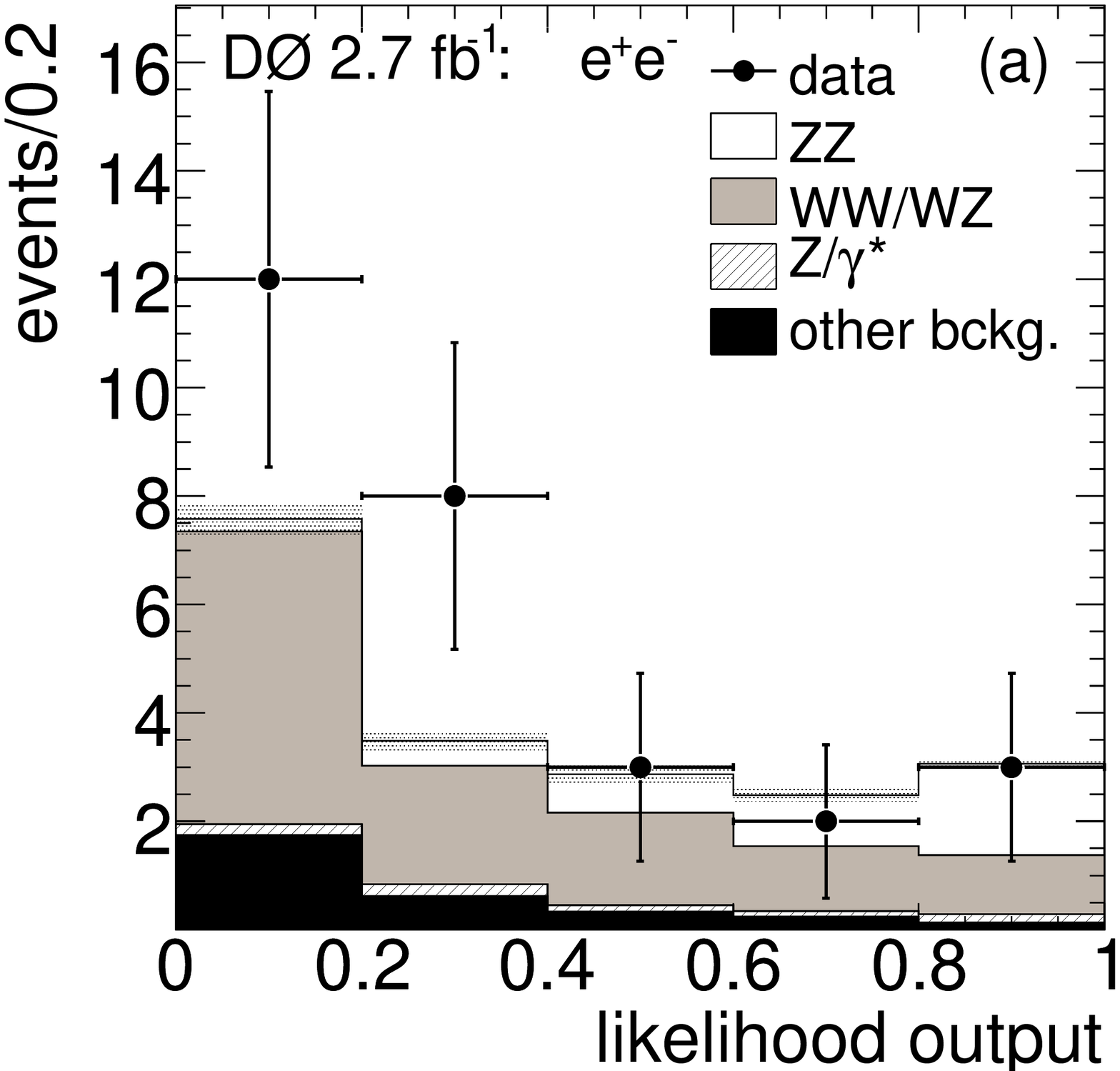}
  \includegraphics[scale=0.2]{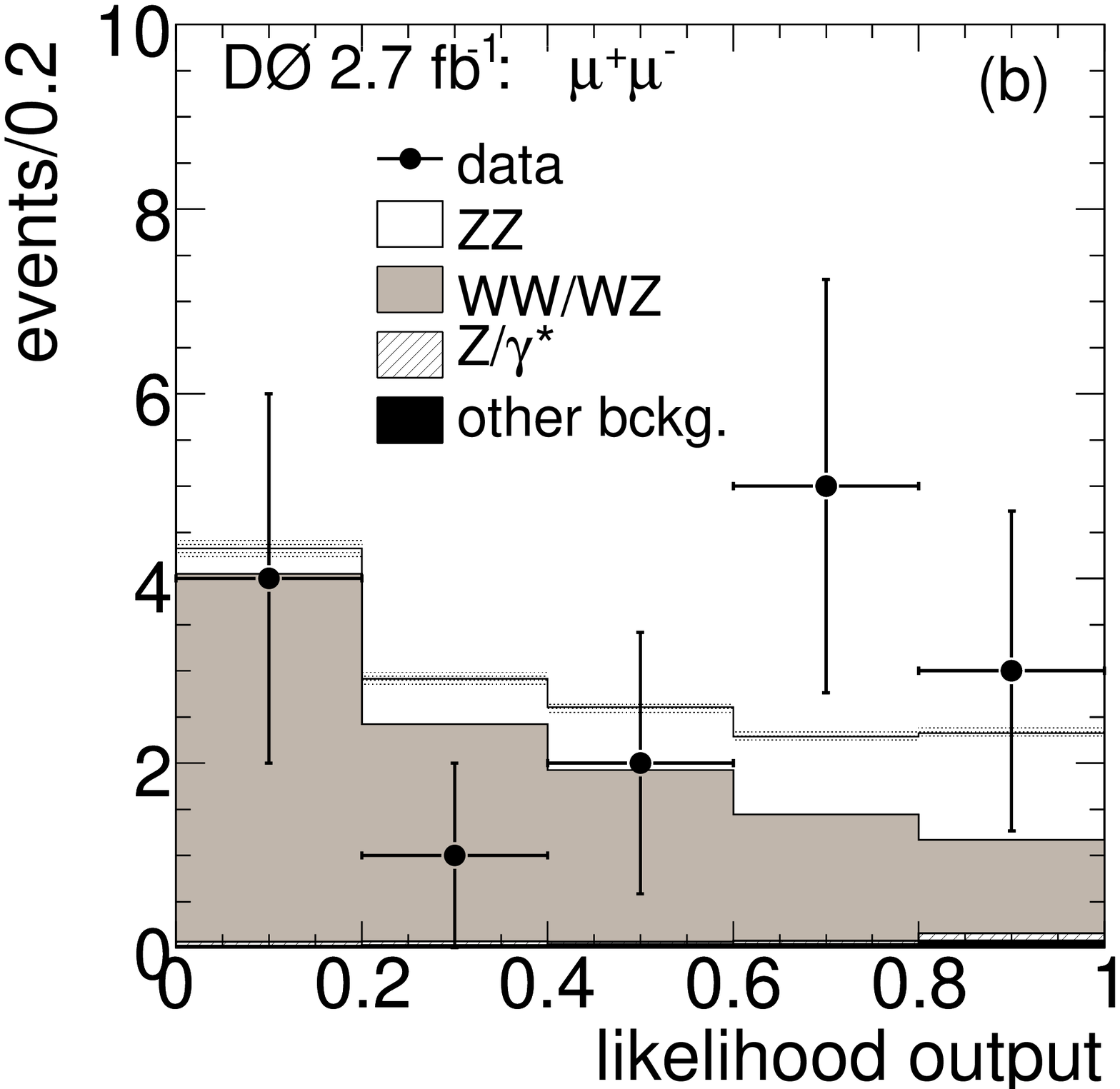}
  \caption{The likelihood distributions for signal and background:
   dielectron events (a) and dimuon events (b). All selection
   requirements have been applied.\label{fig:nll}}
\end{figure}

\begin{table}[hbt]
\caption{Number of predicted dimuon events and yield observed in data after the dilepton
        selection and after the requirement on $\hatmet$.  The uncertainties in the final column are
        statistical only.  If not present, the statistical uncertainty
        is negligible.}
\centering 
\label{tab:cutFlowDimu}
\begin{ruledtabular}
\begin{tabular}{l c c c c}
\hline
Sample  & dilepton selection & $\hatmet$ requirement  \\
\hline
$Z\rightarrow \ell^+ \ell^-$            & $1.30\times10^5$ & ~0.1 $\pm$ 0.1 \\
$Z\rightarrow\tau^+\tau^-$              & 53.3   & 0.09 \hmyOO \\
$W$+Jets                                & --     & $<$ 0.01 \\
$t\bar{t}$                          & 16.0   & 0.21 \hmyOO \\
$WW$                                  & 32.0   & ~9.7 $\pm$ 0.1 \\
$WZ$                                  & 18.3   & 0.82 \hmyOO \\
Predicted Background                & $1.30\times10^5$ & 10.9 $\pm$ 0.3 \\
\hline
$ZZ\rightarrow \ell^+\ell^-\ell'^+\ell'^-$  & 2.89   & 0.00 \hmyOO \\
$ZZ\rightarrow \ell^+\ell^-\nu\bar{\nu}$     & 9.48   & 3.39 \hmyOO \\
Predicted Total                     & $1.30\times10^5$ & 14.3 $\pm$ 0.3 \\
\hline
Data                                & 127,960 & 15 \\
\hline
\end{tabular}
\end{ruledtabular}
\end{table}
The $ZZ$ signal is separated from the remaining backgrounds with significant
$\hatmet$ using a likelihood with the following input variables: the invariant
mass of the dilepton pair, $M_{\ell\ell}$ (for the dielectron channel), the $\chi^{2}$
probability resulting from a refit of the measured lepton momenta under the
constraint that their dilepton mass gives the $Z$ mass (for the dimuon
channel), the transverse momentum of the higher $p_T$ lepton, $\vec{p_{T}}^{(1)}$,
the opening angle between the dilepton pair and the
leading lepton, $\Delta\phi$, and the
cosine of the negative lepton scattering angle in the dilepton rest frame,
$\cos(\theta^{*})$.  Figures~\ref{fig:LLInput_diem}
and~\ref{fig:LLInput_dimu} show the data and predicted distributions of the
variables used in the likelihood for the dielectron and dimuon channels respectively.
Fig.~\ref{fig:nll} shows the likelihood distributions for signal and
backgrounds after all selection requirements.


\section{\label{sec:Systematics}Systematic Uncertainties}
Systematic uncertainties are evaluated separately for the dielectron and the
dimuon samples and for each of the data taking periods.  The
uncertainties affecting the overall scale factor of the MC cross sections are
canceled out by normalizing to the data before the $\hatmet$ cut.
The remaining systematic uncertainties contributing to the
significance and cross section
are dominated by the normalization of the $W$+jets background, the uncertainty
on the $WW$ cross section, the lepton resolution, and the number of $Z$ events
surviving the $\hatmet$ cut.  
The dominant uncertainties are listed in Table~\ref{tab:Systematics}
in which $A_Z$ is the acceptance times efficiency for $Z/\gamma^*\rightarrow\ell^+\ell^-$ 
and $A_{ZZ}$ is the acceptance times efficiency 
for $ZZ/\gamma^{*}\rightarrow \ell^{+}\ell^{-}\nu \bar{\nu}$ where contributions
from $Z/\gamma*\rightarrow\tau^{+}\tau^{-}$ decays are included.
The large uncertainty on the
$W$+jets and remaining $Z$ are due to the uncertainties on the jet to lepton misidentification
rate used in the normalization of the $W$+jets background and the small
statistics available after the $\hatmet$ cut (for both).  Varying
the parameters of the electron and muon smearing in the MC shows that
the effect on the final result is within the statistical uncertainty in almost
all bins.  It is therefore propagated as an uncertainty in the shape of the
likelihood, as are the contributions from jet energy resolution and
the shape of the $ZZ\ p_T$ spectrum.

\section{\label{sec:Xsec}Cross Section and Significance}
A negative log-likelihood ratio (LLR) test statistic is used to
evaluate the significance of the result, taking as input the binned
outputs of the dielectron and dimuon likelihood discriminants
for each of the two data taking periods.
A modified frequentist calculation is used~\cite{b-collie} which
returns the probability of the background only fluctuating to give the
observed yield or higher (p-value) and the corresponding Gaussian
equivalent significance.
The combined dielectron and dimuon channels yield an observed
significance of $2.6$ standard deviations ($2.0$ expected), as reported in Table~\ref{tab:Significance}.

Because of the background normalization method described earlier, 
the measurement of the $ZZ\rightarrow \ell^+\ell^-\nu\bar{\nu}$
production cross section can be therefore expressed in terms of
relative number of events with respect to the $Z\rightarrow \ell^+\ell^-$
sample.

We define a background hypothesis to include the distributions of the
predicted backgrounds shown in Tables~\ref{tab:cutFlowDiem} and~\ref{tab:cutFlowDimu}, and a signal
hypothesis to include these backgrounds and the events from the $ZZ$
process.

To determine the cross section the likelihood distribution in the data
has been fitted allowing the signal normalization to float. 
The scale factor $f$ with respect to the SM cross section
and its uncertainty are determined by the fit.
The  ZZ production cross section is computed by scaling the number of
events predicted by the MC to obtain that in data:
\begin{eqnarray*}
\sigma(ZZ) = \sigma(Z)\frac{A_Z}{A_{ZZ}}\frac{fN^{MC}_{ZZ}}{N_{Z}} \label{eq:xsec} 
\end{eqnarray*}
$A_{ZZ}$ is found to be $4.73 \pm 0.03 \%$ in the dielectron channel and $4.91 \pm 0.03 \%$ in the dimuon channel.
We assume the theoretical cross section for  $Z/\gamma^*\rightarrow
\ell^+\ell^-$ in the mass window $60 < M_{\ell\ell} < 130$~GeV:
$\sigma(Z) = 256.6^{+5.1}_{-12.0}$~pb~\cite{b-zxsec,b-zxsec2}.
Using the ratio of  the $ZZ$ to $ZZ/\gamma^*$ cross sections
computed with MCFM \cite{campbell-ellis} at NLO, we scale the $ZZ/\gamma^*$ cross section
down by 3.4\% to give a pure $ZZ$ cross section.
The resulting cross section for $\ppbar \rightarrow ZZ+X$ is
\begin{eqnarray*}
\sigma(ZZ) = 2.01 \pm 0.93 (stat.)\pm 0.29 (sys.)\ \mathrm{pb}.
\end{eqnarray*}
This can be compared with the predicted SM cross section
of $1.4\pm 0.1$~pb~\cite{campbell-ellis} at $\sqrt{s} = 1.96$~TeV.

\begin{table}[hbt]
\caption{The values assumed by the dominant systematic uncertainties for the various Monte
	Carlo signal and background samples in	the dielectron and
	dimuon channels.}
\label{tab:Systematics}
\begin{ruledtabular}
\begin{tabular}{l c c}
\hline
Systematic uncertainty & dielectron & dimuon\\
 & (\%) & (\%)\\
\hline
$W$+Jets normalization & 16 & -- \\
$WW$ and $WZ$ & \\
 Theoretical cross sections & 7 & 7\\
Number of $Z$ events surviving  & \\
the $\hatmet$ cut & 18 & 3\\
\hline
\hline
Systematic uncertainty on the & \multicolumn{2}{c}{Uncertainty} \\
cross section & \multicolumn{2}{c}{(\%)} \\
\hline
$Z/\gamma^{*}\rightarrow\ell^+\ell^-$ & \\
 theoretical cross section & $^{+2.0}_{-5.0}$ & $^{+2.0}_{-5.0}$\\
${A_Z}/{A_{ZZ}}$ ratio from pdf \\
uncertainties & 1.8 & 1.8\\
${A_Z}/{A_{ZZ}}$ ratio from modeling & \\
of the veto efficiency & 0.8 & 0.8\\
${A_Z}/{A_{ZZ}}$ ratio from modeling & \\
of the $ZZ$ $p_{T}$ spectrum  & 3.0 & 3.0\\
\end{tabular}
\end{ruledtabular}
\end{table}

\begin{table}
\caption{Estimated significance for background only to fluctuate
  to at least the observed yield for the combined dielectron and
  dimuon channels in the two data taking periods.}
\label{tab:Significance}
\begin{ruledtabular}
\begin{tabular}{l c c}
\hline
& expected ($\sigma$) & observed ($\sigma$) \\
\hline
p-value & 0.0244 & 0.0042\\
significance & 2.0 & 2.6 \\
\hline
\end{tabular}
\end{ruledtabular}
\end{table}

\section{\label{sec:Conclusion}Conclusion}

We performed a measurement of the production cross section of $ZZ\rightarrow
\ell^+\ell^-\nu\bar{\nu}$ using 2.7 fb$^{-1}$ of data collected by the D0
experiment at a center of mass energy of 1.96~TeV. We observe a signal with a
$2.6$ standard deviations significance ($2.0$ expected) and measure a
cross section
$\sigma(p\bar{p}\to ZZ)=2.01\pm 0.93 \mathrm{(stat.)}\pm 0.29 \mathrm{(sys.)}$~pb.
This is in agreement with the standard model prediction of $1.4$~pb~\cite{campbell-ellis}.

%
We thank the staffs at Fermilab and collaborating institutions, 
and acknowledge support from the 
DOE and NSF (USA);
CEA and CNRS/IN2P3 (France);
FASI, Rosatom and RFBR (Russia);
CNPq, FAPERJ, FAPESP and FUNDUNESP (Brazil);
DAE and DST (India);
Colciencias (Colombia);
CONACyT (Mexico);
KRF and KOSEF (Korea);
CONICET and UBACyT (Argentina);
FOM (The Netherlands);
STFC (United Kingdom);
MSMT and GACR (Czech Republic);
CRC Program, CFI, NSERC and WestGrid Project (Canada);
BMBF and DFG (Germany);
SFI (Ireland);
The Swedish Research Council (Sweden);
CAS and CNSF (China);
Alexander von Humboldt Foundation (Germany);
and the
Istituto Nazionale di Fisica Nucleare (Italy).
%

\end{document}